\documentstyle[12pt,epsfig,psfig]{article}
\begin{document}

\begin{center}
{\Large \bf CsI(Tl) for WIMP dark matter searches}

\baselineskip=24pt
\vspace{0.5cm}
{\large V. A. Kudryavtsev, N. J. C. Spooner, D. R. Tovey, 
J. W. Roberts, \\
M. J. Lehner, J. E. McMillan, 
P. K. Lightfoot, T. B. Lawson, C. D. Peak, 
R. L\"uscher}

\vspace{0.2cm}
{\it Department of Physics and Astronomy, University of Sheffield, 
Sheffield S3 7RH, UK}

\vspace{0.3cm}
{\large J. C. Barton}

\vspace{0.2cm}
{\it Birkbeck College, London WC1E 7HX, UK}

\vspace{0.5cm}
{\large \bf Abstract}
\end{center}
\vspace{0.2cm}

We report a study of CsI(Tl) scintillator to assess its applicability 
in experiments to search for dark matter particles. 
Measurements of the mean scintillation pulse shapes due to nuclear and 
electron recoils have been performed. We find that, as with NaI(Tl), 
pulse shape analysis can be used to 
discriminate between electron and nuclear recoils down to 4 keV. 
However, the discrimination factor is 
typically (10-15)\% better than in NaI(Tl) above 4 keV. 
The quenching factor for caesium and iodine 
recoils was measured and found to increase from 11\% to $\sim17\%$ 
with decreasing recoil energy from 60 
to 12 keV.  Based on these results, the potential sensitivity of 
CsI(Tl) to dark matter particles in the 
form of neutralinos was calculated. We find an improvement over NaI(Tl) 
for the spin independent 
WIMP-nucleon interactions up to a factor of 5 assuming comparable electron 
background levels in the two scintillators.

\vspace{0.3cm}


\vspace{0.2cm}
\noindent Corresponding author: V. A. Kudryavtsev, Department of Physics 
and Astronomy, University of Sheffield, Hicks Building, Hounsfield Rd., 
Sheffield S3 7RH, UK

\noindent Tel: +44 (0)114 2224531; \hspace{2cm} Fax: +44 (0)114 2728079; 

\noindent E-mail: v.kudryavtsev@sheffield.ac.uk

\pagebreak

\noindent {\large \bf 1. Introduction}
\vspace{0.3cm}

NaI(Tl) crystals are widely used in searches for Weakly Interacting 
Massive Particles (WIMPs) 
- possible constituents of the galactic dark matter 
(see for example \cite{UKDMC,DAMA,Gerbier,Fushimi}). However, a particular 
disadvantage of such crystals is their hygroscopic nature.  
This results in handling problems, in 
particular degradation of the surfaces through contact with moisture 
(such as from air humidity), which 
in turn increases the potential for generating atypical events from 
background electron or alpha 
interactions at the crystal surfaces \cite{Gerbier,VK,Nigel}. 
CsI(Tl) crystals are much less hygroscopic than NaI(Tl) and 
hence offer a potential alternative with these surface complications reduced. 

In this paper we report 
studies of the properties of CsI(Tl) relevant to its use as a possible 
detector for dark matter. 
To compete, or improve on, the sensitivity of NaI(Tl) to WIMP 
interactions CsI(Tl) needs to 
satisfy several requirements: i) it should provide discrimination 
between electron and nuclear recoils 
down to low energies (2-4 keV) using pulse shape analysis, 
ii) the quenching factor of caesium and 
iodine recoils should not be less than 10\%, and 
iii) the background rate should be comparable to that 
achieved in low background NaI(Tl) crystals 
(1-2 dru, differential rate unit, 1 dru = 1 
event/kg/day/keV). Our study here has concentrated on requirements 
i) and ii) since requirement iii), 
the background rate, is not an intrinsic property of CsI(Tl) but 
rather will vary from crystal to crystal 
depending on the selected starting materials, purification and 
manufacturing processes used (see 
Section 5).  Recently, similar studies have been performed by Kim et al. 
\cite{Kim} and P\'ecourt et al. \cite{Pecourt}. 
P\'ecourt et al. \cite{Pecourt} used a high energy neutron beam to 
measure quenching factors of nuclear recoils in 
CsI(Tl). Results indicate an increase in scintillation efficiency with 
decreasing recoil energy down to 25 
keV suggesting CsI(Tl) detectors may be more favourable for dark matter 
searches than had previously 
been realised. They also measured distributions of mean scintillation 
pulse decay times induced by 
electron and nuclear recoils down to 5 keV and found better discrimination 
compared to NaI(Tl) 
crystals. In this work we have extended measurements of quenching factor 
down to recoil energies of 
12 keV and have made detailed studies of the pulse shape and of the 
distribution of decay time 
constants due to electron and neutron induced recoils down to 2 keV. 
Finally, based on results of these 
studies we have estimated the discrimination efficiency achievable 
with CsI(Tl) and hence the 
sensitivity of CsI(Tl) to the spin independent WIMP-nucleon and spin 
dependent WIMP-proton 
interactions assuming various intrinsic background rates.

\vspace{0.5cm}
\noindent {\large \bf 2. Mean pulse shapes}
\vspace{0.3cm}

Measurements of the scintillation pulse characteristics of CsI(Tl) 
were achieved using a crystal of 
3 inch diameter $\times$ 1.5 inch close coupled to two 3 inch ETL 9265KB 
photomultiplier tubes (PMTs). 
The crystal was polished and its curved surface wrapped in PTFE to provide 
a reflective coating. A 
light yield of about 5.5 photoelectrons (p.e.) / keV was achieved with 
this detector. Note that the choice 
of PMT was made here with respect to low background characteristics - 
the 9265KB (bialkali 
photocathode) being our standard choice in NaI(Tl) dark matter experiments. 
PMTs with optimum 
spectral response matched to the CsI(Tl) peak emission at 550 nm 
\cite{Grassman} would have higher quantum 
efficiency at this wavelength ($\sim25\%$ compared with $\sim10\%$ for 
bialkali) but are not available with known 
low background characteristics. Three different radioactive sources 
were used in the pulse shape 
experiments: a $^{57}$Co gamma source (122 keV) for energy calibration; 
a $^{60}$Co source to provide 
populations of low energy Compton scatter electrons; and a $^{252}$Cf 
neutron source to produce nuclear 
recoil events. The energy resolution was found to be 7\% at 122 keV, 
which exceeds the `statistical 
resolution' calculated for a light yield of 5.5 p.e./keV, 
by a factor of 2 only. The measurements were 
performed mainly at a controlled temperature of $(10.0\pm0.5)^{\circ}$ C, to 
reduce PMT noise, and at room 
temperature $(23\pm1)^{\circ}$ C for comparison. 

Mean pulse shapes for electron- and nuclear recoil- induced events 
for measured energies of 5-6 
keV are shown in Figure 1. The pulse shape for nuclear recoil events 
is scaled down by a factor of 10 
to avoid overlap of the two distributions. The energy range of 5-6 keV 
is chosen here because only at 
low energies do the nuclear recoils induced by neutrons from $^{252}$Cf 
dominate over electron recoils 
induced by the accompanying gammas (the fraction of electron pulses 
does not exceed 2\% at 5-6 keV). 
Two prominent decay components (fast and slow) are clearly seen in both 
distributions shown in 
Figure 1. This agrees with early measurements performed at high energies 
(see \cite{Birks,Valentine,Schotanus,Benrachi} and references 
therein). The mean pulse shapes were found to be well fitted by the sum 
of two exponential 
components with time constants at $10^{\circ}$ C of 
$\tau_{g1}=(860\pm60)$ ns, $\tau_{g2}=(2600\pm140)$ ns for gamma-induced 
pulses and $\tau_{n1}=(750\pm20)$ ns, $\tau_{n2}=(1800\pm100)$ ns for 
neutron-induced pulses. 

\vspace{0.5cm}
\noindent {\large \bf 3. Pulse shape discrimination}
\vspace{0.3cm}

In further tests we have applied the standard procedure of pulse shape 
analysis adopted by the 
UK Dark Matter Collaboration for NaI(Tl) dark matter detectors 
\cite{UKDMC,VK,Nigel,Dan}. 
Pulses from both PMTs 
were integrated using a buffer circuit and then digitised using a 
LeCroy 9350A oscilloscope driven by 
a Macintosh computer running Labview-based data acquisition software. 
The digitised pulse shapes 
were passed to the computer and stored on disk. Final analysis was 
performed on the sum of the pulses from the two PMTs. 

Our standard procedure of data analysis, described in \cite{UKDMC}, 
involves the fitting of a single 
exponential to each integrated pulse to obtain the index of the exponent, 
$\tau$. Although scintillation pulses 
from CsI(Tl) have an additional second component (see Section 2), 
the pulses can nevertheless be well 
fitted by a single exponential if fits are restricted to data below 
1500 ns (see Figure 1). This fraction of 
the pulse contains the major contribution to the integrated pulse 
amplitude so that the distortion of the 
fit due to the presence of the second exponential at large time scales 
was found to be insignificant.  
This approximation has the advantage that a three parameter fit can be 
used on each pulse and a simple 
discrimination parameter defined, rather than a considerably more 
complicated six parameter fit in the 
case of two decay constants. The free fit parameters used are: 
the time constant of the single exponent, 
$\tau$; a normalisation constant and the start time of the pulse.

For each experiment (with gamma or neutron source) the distribution of the 
number of events 
versus the time constant of the exponent ($\tau$) was generated for a 
range of energy bins. $\tau$-Distributions 
for each population of pulses can be approximated by a gaussian in 
$\ln(\tau)$ \cite{UKDMC,VK} (for a more detailed 
discussion of the distributions see \cite{Dan} and references therein):

\begin{equation}
{{dN}\over{d\tau}} = {{N_o}\over{\tau \sqrt{2\pi} \ln w}} \cdot
\exp \Big[ {{-(\ln \tau-\ln \tau_o)^2}
\over{2(\ln w)^2}} \Big]
\end{equation}

\vspace{0.5cm}
The $\tau$-distributions have been fitted with a gaussian in $\ln(\tau)$ 
with several free parameters as 
follows. For distributions from the $^{60}$Co gamma source a 
3-parameter fit was used with free parameters 
$\tau_o$, $w$ and $N_o$. In the experiments with the $^{252}$Cf neutron 
source both neutrons and gammas (from the 
source as well as from local radioactivity) are detected. The resulting 
$\tau$-distribution is thus fitted with 
two gaussians. However, the parameters $\tau_o$ and $w$ for events 
initiated by Compton electrons are known 
from experiments with the $^{60}$Co gamma source. Assuming the value of 
$w$ (called the width parameter) 
for the nuclear recoil distribution is the same as that of the gamma 
distribution (since the width is 
determined mainly by the number of photoelectrons) again a 3-parameter 
fit can be applied \cite{VK}. In this 
case the free parameters are the number of neutrons, $N_{on}$, 
number of gammas, $N_{og}$, and the mean value 
of the exponent for the neutron distribution, $\tau_{on}$. 

Examples of $\tau$-distributions of electron and nuclear recoils 
together with the gaussian fits are 
shown in Figure 2 for measured energies 8-9 keV. Note again, that 
both nuclear and electron recoil 
pulses are present in the $\tau$-distribution obtained with the neutron 
source (see also \cite{VK}) (the contribution 
of electron recoils is shown by dashed line).

The mean values $\tau_o$ of log-gaussian fits as a function of measured 
energy are shown in Figure 3 
for electron and nuclear recoils and for two temperatures of the crystal, 
$23^{\circ}$ C (room temperature) and 
$10^{\circ}$ C. There is a strong energy and temperature dependence of the 
parameter $\tau_o$. However, the fact 
that the values of $\tau_o$ increase with decreasing temperature is found 
not to influence the discrimination 
(see below) since the ratio of nuclear to electron recoil time constants 
and the width parameter remain almost constant. 

The ratio of nuclear to electron recoil time constants as a function of 
energy for $10^{\circ}$ C is shown 
in Figure 4 in comparison with values for NaI(Tl) \cite{NAIAD}. 
The lower ratio of time constants found in the 
CsI(Tl) crystal results in better discrimination between nuclear 
and electron recoils than in NaI(Tl). The 
real gain is reduced due to the larger value of the width $w$ for 
CsI(Tl), shown in Figure 5. No 
discrimination was found below 4 keV, which is true also for NaI(Tl) crystals. 

The ratio of time constants at $10^{\circ}$ C is found empirically to be 
approximated by: $R_{\tau}$ ($E > 4$ keV)$ = 
a + b \cdot \exp [ ( c - E ) / d ]$ and $R_{\tau}$ ($E < 4$ keV) = 1, 
where the parameters are: $a=0.65$, $b=0.35$, $c=3$ keV 
and $d=4.5$ keV for the CsI(Tl) crystal and $a=0.75$, $b=0.25$, $c=3$ keV 
and $d=5$ keV for NaI(Tl) \cite{NAIAD}. The 
parameterisations are shown in Figure 3 by solid lines. The width $w$ 
of the $\tau$-distribution is a function 
of the number of photoelectrons and can be parameterised as: 
$w = 1 + f \cdot ( N_{pe} )^{-1/2}$, where $N_{pe}$ is the 
number of photoelectrons and parameter $f=2.4$ for CsI(Tl) at $10^{\circ}$ C 
and $f=1.7$ for NaI(Tl) at $10^{\circ}$ C \cite{NAIAD}. 
Note that the dependence of parameter $f$ on the crystal and its 
temperature was found to be insignificant for NaI(Tl) crystals.

To quantify the discrimination between nuclear and electron recoils 
we define a discrimination 
factor $D = 1 - S$, where $S$ is the fraction of the electron and nuclear 
recoil $\tau$-distributions that are 
overlapping - the distributions being first normalised to a total count of 
1 \cite{NAIAD}.  For fully overlapping 
distributions, $R_{\tau} = 1$ and $D = 0$ and there is no discrimination. 
For separated distributions $D = 1$ and 
there is full event by event discrimination. The discrimination factors 
as a function of measured energy 
are shown in Figure 6 for CsI(Tl) and NaI(Tl) detectors with different 
light yields derived using the 
aforementioned parameterisations of $R_{\tau}$ and $w$. It can be seen 
from Figure 6 that for a given light yield, 
above $\sim4$ keV, the discrimination power of CsI(Tl) is always better 
than that of NaI(Tl), typically by $\sim(10-15)\%$.

\vspace{0.5cm}
\noindent {\large \bf 4. Quenching factor}
\vspace{0.3cm}

The measurement of scintillation efficiency or quenching factor of 
caesium and iodine recoils in 
CsI(Tl) was performed using the 2.85 MeV neutron beam facility of 
Sheffield University with 1 inch 
diameter $\times$ 1 inch CsI(Tl) crystal. The experimental set up and analysis 
procedure adopted was similar 
to that described in \cite{Dan,Dan1}. Neutrons were scattered through angles 
of (45--120) degrees with respect 
to the primary flux, yielding nuclear recoil energies of 12.5--65 keV. 
As the masses of caesium and 
iodine nuclei are very similar, discrimination between them was not 
possible so the results obtained refer to both types of nuclei. 

The measured quenching factor $Q = E_m / E_r$, where $E_m$ is measured 
energy and $E_r$ is recoil energy, 
is shown in Figure 7a as a function of recoil energy together with 
the measurement reported in \cite{Pecourt}. 
There is a good agreement between the two measurements. Both measurements 
indicate an increase in 
quenching factor with decreasing recoil energy. This important result 
significantly affects the WIMP 
sensitivity of CsI(Tl) crystals (see Section 5), in particular to spin 
independent interactions for which 
the recoil spectrum falls most rapidly with energy. To calculate the 
sensitivities to WIMP interactions 
the dependence of the quenching factor on the recoil energy was 
converted to measured energy versus 
recoil energy. This is shown in Figure 7b together with our 
empirical parameterisation. 

\vspace{0.5cm}
\noindent {\large \bf 5. Sensitivity of CsI(Tl) detectors to WIMP search}
\vspace{0.3cm}

Taking the results of the measurements reported above, we can 
calculate the potential sensitivity 
of CsI(Tl) to WIMP-proton spin dependent and WIMP-nucleon spin 
independent interactions. 
Another parameter, which is important in such calculations, 
is the background rate. As mentioned in 
Section 1, this is an uncertain parameter, since the background rate 
(for instance, if determined by 
$^{137}$Cs, $^{135}$Cs and $^{134}$Cs contamination) is a property 
of the particular crystal used and may vary by an 
unknown factor from one crystal to another. Therefore, in our 
simulations we have chosen, for 
illustrative purposes, to use two values of background rate: 
2 dru and 10 dru. The first one is the 
background rate reached currently in low-background NaI(Tl) 
detectors used to search for dark matter 
particles in the form of neutralinos \cite{DAMA,Gerbier,NAIAD}. 
The second one is a more realistic estimate of the 
background in the CsI(Tl) contaminated with radioactive isotopes 
of $^{137}$Cs, $^{135}$Cs and $^{134}$Cs. We have 
assumed in the simulations that the energy threshold was 2 keV, 
the light yield is 3 p.e./keV, the 
discrimination between nuclear and electron recoils starts at 
4 keV and the parameters of the log-gaussian fits are as described above. 
We have used our empirical parameterisation for the relation 
between measured energy and recoil energy, shown in Figure 7b. 
The spin factor of $^{133}$Cs has been 
calculated using the odd group model \cite{David}. The spin-dependent form 
factor and spin factor for iodine 
have been taken from \cite{David,RD,LS} and the spin dependent form factor 
for caesium has been assumed to be 
the same as for iodine. A higgsino-type neutralino has been assumed. 
The form factors for caesium 
and iodine for spin independent interactions have been calculated 
using Helm's parameterisation \cite{Helm} 
(see also the discussion in \cite{LS}). An $A^2$-enhancement of the spin 
independent cross-section has been 
assumed. The halo parameters have been taken as follows: local 
dark matter density $\rho_{dm}$ = 0.3 
GeV/cm$^3$, the Maxwellian dark matter velocity distribution with 
$v_o$ = 220 km/s, the local Galactic escape 
velocity $v_{esc}$ = 650 km/s and the Earth's velocity relative to the 
dark matter distribution $v_{Earth}$ = 232 km/s.

The resulting sensitivity plots are presented in Figures 8a and 8b 
for spin dependent and spin 
independent interactions, respectively. It can be seen that no 
improvement in sensitivity is expected 
from CsI(Tl) compared with NaI(Tl) for spin-dependent WIMP-proton 
interactions even when the 
lower background rate is assumed. This is due to the step nature 
of the form factor functions of both 
caesium and iodine, while that of sodium is smoother resulting 
in a dominant contribution from sodium 
interactions in NaI(Tl) detectors. (Note that more precise 
calculations of the caesium spin factor may 
modify the sensitivity curve but are unlikely to change the conclusion). 
However, it is found that 
CsI(Tl) (even with higher background rate) is more sensitive to 
spin independent interactions than 
NaI(Tl) at a given light yield. This is because both caesium 
and iodine nuclei have large mass and 
hence the WIMP-nucleus interaction rate is affected by a 
large $A^2$-enhancement factor. The increase in 
nuclear recoil scintillation efficiency with decreasing energy 
also contributes to the improvement in the sensitivity.

\vspace{0.5cm}
\noindent {\large \bf 6. Conclusions}
\vspace{0.3cm}

Measurements of various characteristics of CsI(Tl) relevant to dark 
matter searches have been 
performed in the range of visible energies 2-20 keV - important for 
WIMP-nucleus interactions. 
Discrimination between nuclear and electron recoils was found to 
be typically (10-15)\% better than in 
NaI(Tl) crystals for a given light yield and the scintillation 
efficiency factor for caesium and iodine 
recoils was found to increase with decreasing recoil energy, 
in good agreement with the results reported 
in \cite{Pecourt}. Based on these results the sensitivity of CsI(Tl) 
to spin independent WIMP-nucleon 
interactions appears to be better than that of NaI(Tl) up to a 
factor of 5 for a given light yield assuming 
that the background rate achievable in CsI(Tl) can be similar 
to that found in low background NaI(Tl) 
crystals. For spin dependent WIMP-proton interactions the 
sensitivity of CsI(Tl) appears to be worse 
than that of NaI(Tl).

Based on these conclusions it appears that CsI(Tl) could provide a 
dark matter target with similar 
or improved sensitivity over NaI(Tl) but with the additional 
advantage of easier control over potential 
surface background events. However, there remains the need to 
assess the intrinsic electron 
background in CsI(Tl) in particular that due to $^{137}$Cs. This, 
and investigation of the application of 
radiopurification techniques used in NaI to CsI, will be the 
subject of future studies.  

\vspace{0.5cm}
\noindent {\large \bf 7. Acknowledgements}
\vspace{0.3cm}

The authors wish to thank the following for their support: 
PPARC, Electron Tubes Ltd. (J.E.M.), 
Royal Society (R.L., ESEP grant N 81RS-57440). 
We are grateful to D. Lewin (RAL) for useful discussions.

\pagebreak

{\large \bf Figure captions}
\vspace{0.3cm}

\noindent Figure 1. Mean pulse shapes of nuclear and electron recoil pulses 
in CsI(Tl) crystal for measured 
energy 5-6 keV. Pulse shape of nuclear recoils is scaled down 
by a factor of 10. Fits with two 
exponential components to the measured distributions are 
shown by solid curves.

\vspace{0.3cm}
\noindent Figure 2. Time constant distributions of nuclear and 
electron recoil pulses: a) $\tau$-dist\-ribution measured 
with gamma source; b) $\tau$-distribution measured with neutron source. 
Nuclear and electron recoil pulses 
are present in b). Fits with one (a) or two (b) 
log-gaussian components are shown by solid curves. 
Contribution of gamma-induced pulses is shown by dashed curve in b).

\vspace{0.3cm}
\noindent Figure 3. Mean time constants of log-gaussian 
distributions versus measured energy are shown for 
nuclear and electron recoils at different temperatures: crosses -- 
nuclear recoils at room temperature; 
squares -- electron recoils at room temperature; triangles -- nuclear 
recoils at 10$^{\circ}$ C; diamonds -- electron recoils at 10$^{\circ}$ C.

\vspace{0.3cm}
\noindent Figure 4. Ratio of nuclear to electron mean time constant 
for CsI(Tl) (diamonds) and NaI(Tl) 
(triangles) crystals at 10$^{\circ}$ C together with empirical parameterisations.

\vspace{0.3cm}
\noindent Figure 5. Width of log-gaussian distributions as a 
function of measured energy for CsI(Tl) 
(diamonds) and NaI(Tl) (triangles) crystals at 10$^{\circ}$ C together 
with empirical parameterisations.

\vspace{0.3cm}
\noindent Figure 6. Discrimination factor as a function of measured energy 
for CsI(Tl) (squares -- light yield = 3 
p.e./keV, triangles -- light yield = 8 p.e./keV) and NaI(Tl) 
(diamonds -- light yield = 3 p.e./keV, crosses -- 
light yield = 8 p.e./keV) from \cite{NAIAD} (see text for details).

\vspace{0.3cm}
\noindent Figure 7. a) Quenching factor of nuclear recoils in 
CsI(Tl) crystal as a function of recoil energy 
(squares) is shown together with the results from \cite{Pecourt} (triangles). 
b) Visible energy versus recoil 
energy: squares -- present data, triangles -- data from \cite{Pecourt}, solid 
line -­ empirical parameterisation.

\vspace{0.3cm}
\noindent Figure 8. Estimates of the sensitivity of CsI(Tl) 
detector with 10 kg$\times$years exposure to a) spin dependent WIMP-proton and 
b) spin independent WIMP-nucleon interactions compared with NaI(Tl) 
detector \cite{NAIAD}: a) solid curve ­- 
our estimates for CsI(Tl) crystal with a background rate of 2 dru 
and light yield 3 p.e./keV, dashed 
curve ­- UKDMC limits from \cite{UKDMC} recalculated \cite{NAIAD} with new halo 
parameters and spin factors, dotted 
curve ­- estimates for NaI(Tl) crystal \cite{NAIAD} for light yield 3 p.e./keV; 
b) solid curve ­- our estimates for 
CsI(Tl) crystal with a background rate of 2 dru and light yield 3 p.e./keV, 
dash-dotted curve -- our 
estimates for CsI(Tl) crystal with a background rate of 10 dru 
and light yield 3 p.e./keV, dashed curve 
­- UKDMC limits from \cite{UKDMC} recalculated \cite{NAIAD} with new 
halo parameters and spin factors, dotted curve ­- 
estimates for NaI(Tl) crystal \cite{NAIAD} for light yield 3 p.e./keV; 
(see also text for details).

\pagebreak

\begin{figure}[htb]
\begin{center}
\epsfig{figure=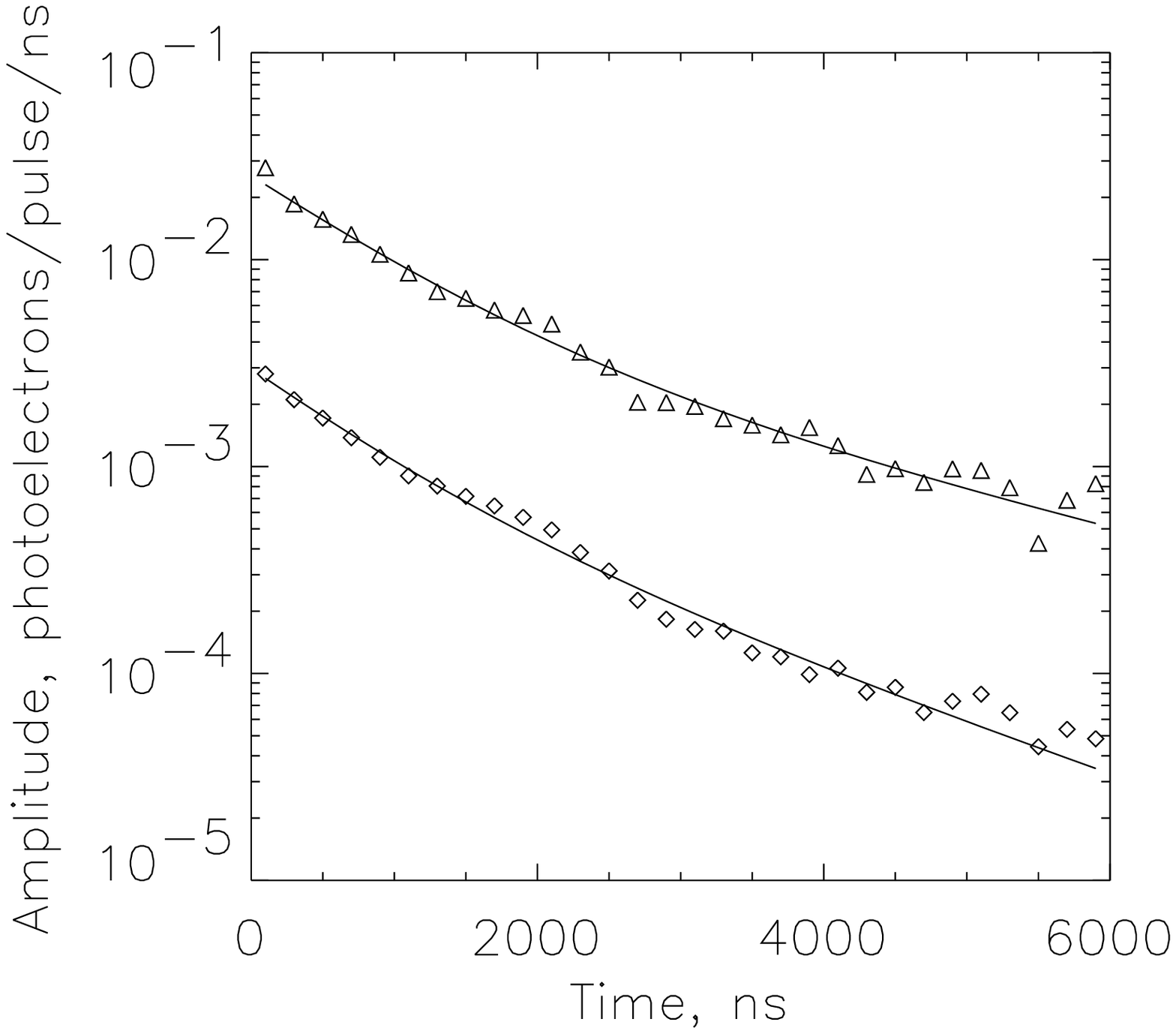,height=14cm}
\caption { }
\end{center}
\end{figure}

\pagebreak

\begin{figure}[htb]
\begin{center}
\epsfig{figure=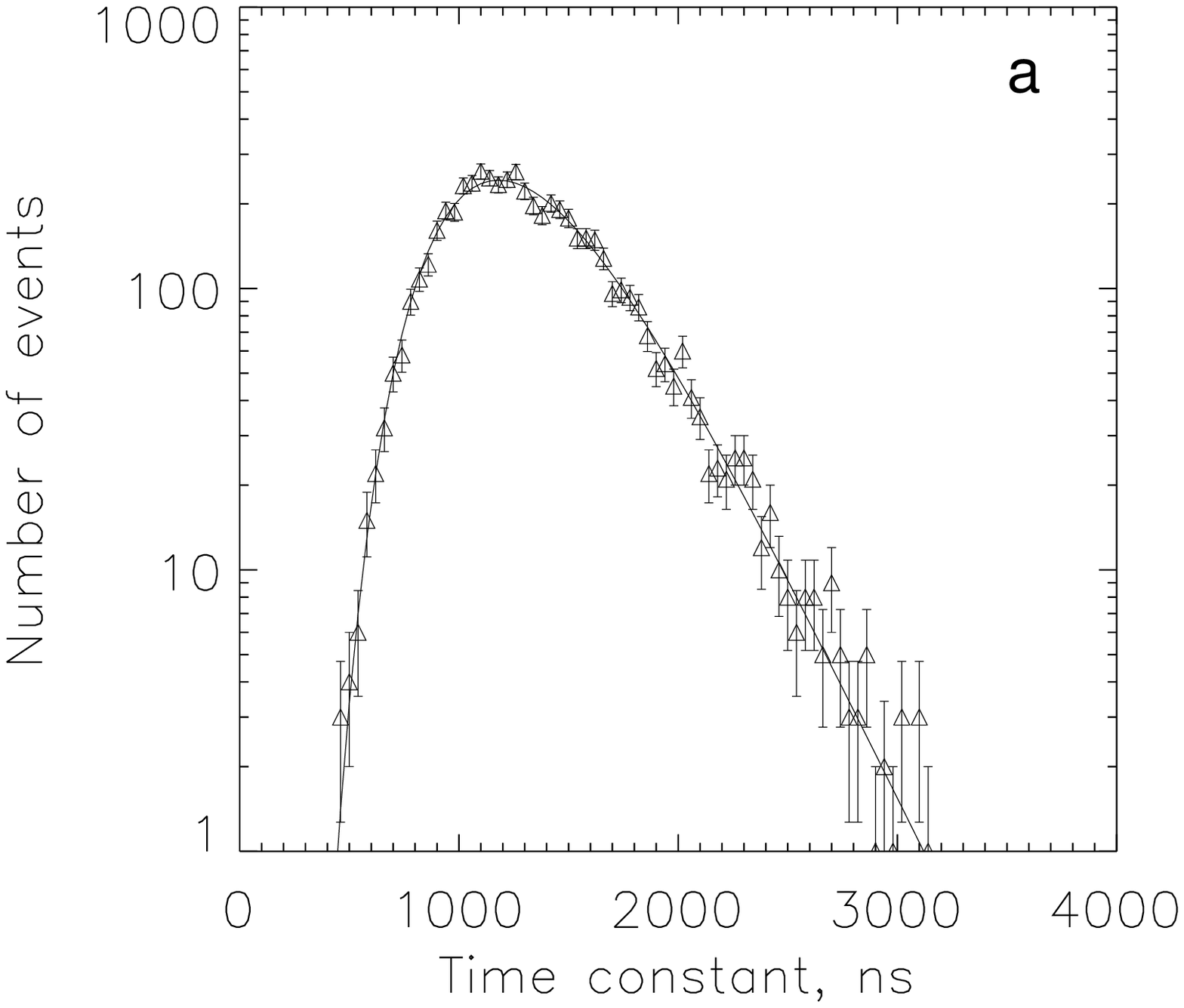,height=8cm}
\epsfig{figure=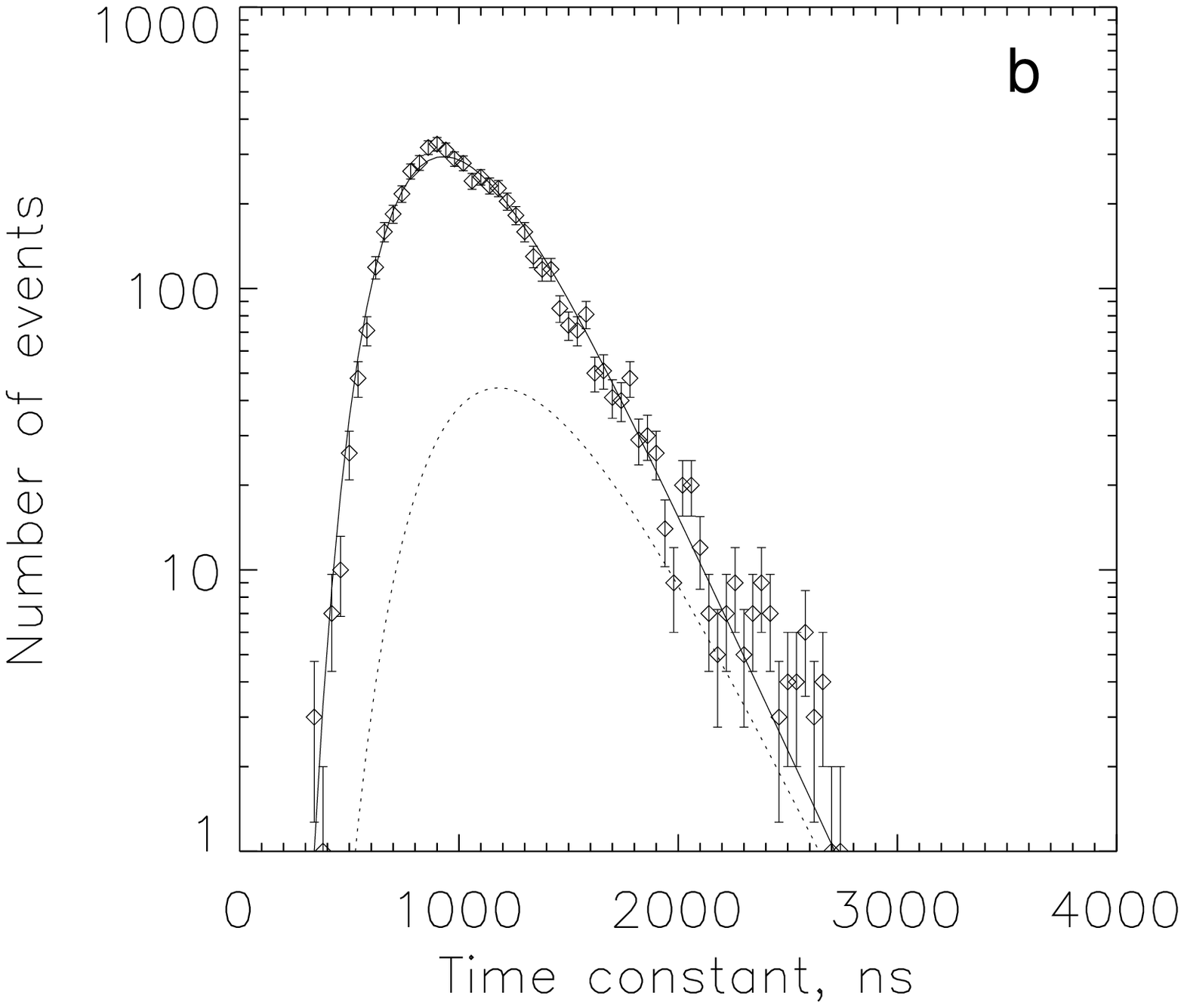,height=8cm}
\caption { }
\end{center}
\end{figure}

\pagebreak

\begin{figure}[htb]
\begin{center}
\epsfig{figure=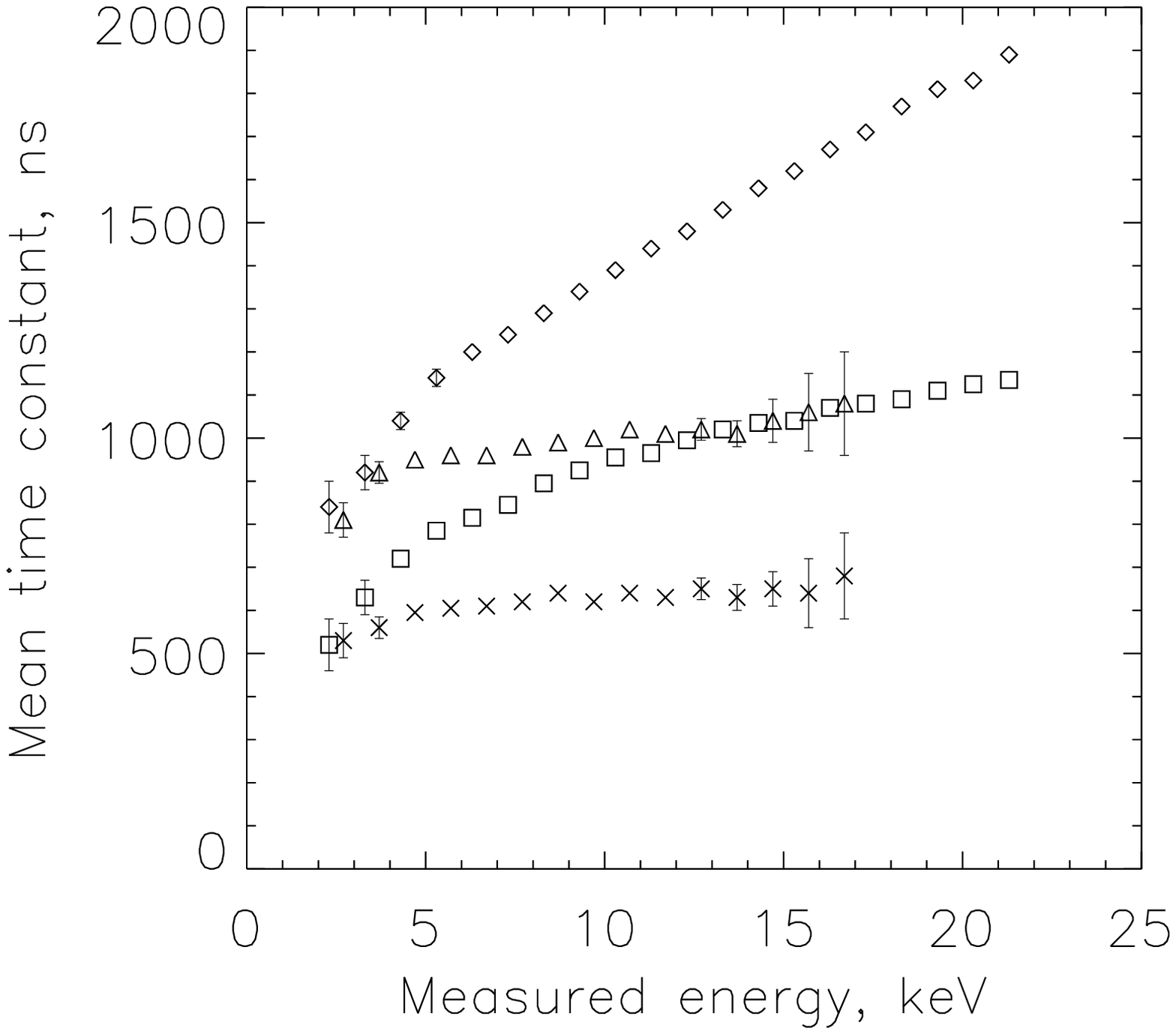,height=14cm}
\caption { }
\end{center}
\end{figure}

\pagebreak

\begin{figure}[htb]
\begin{center}
\epsfig{figure=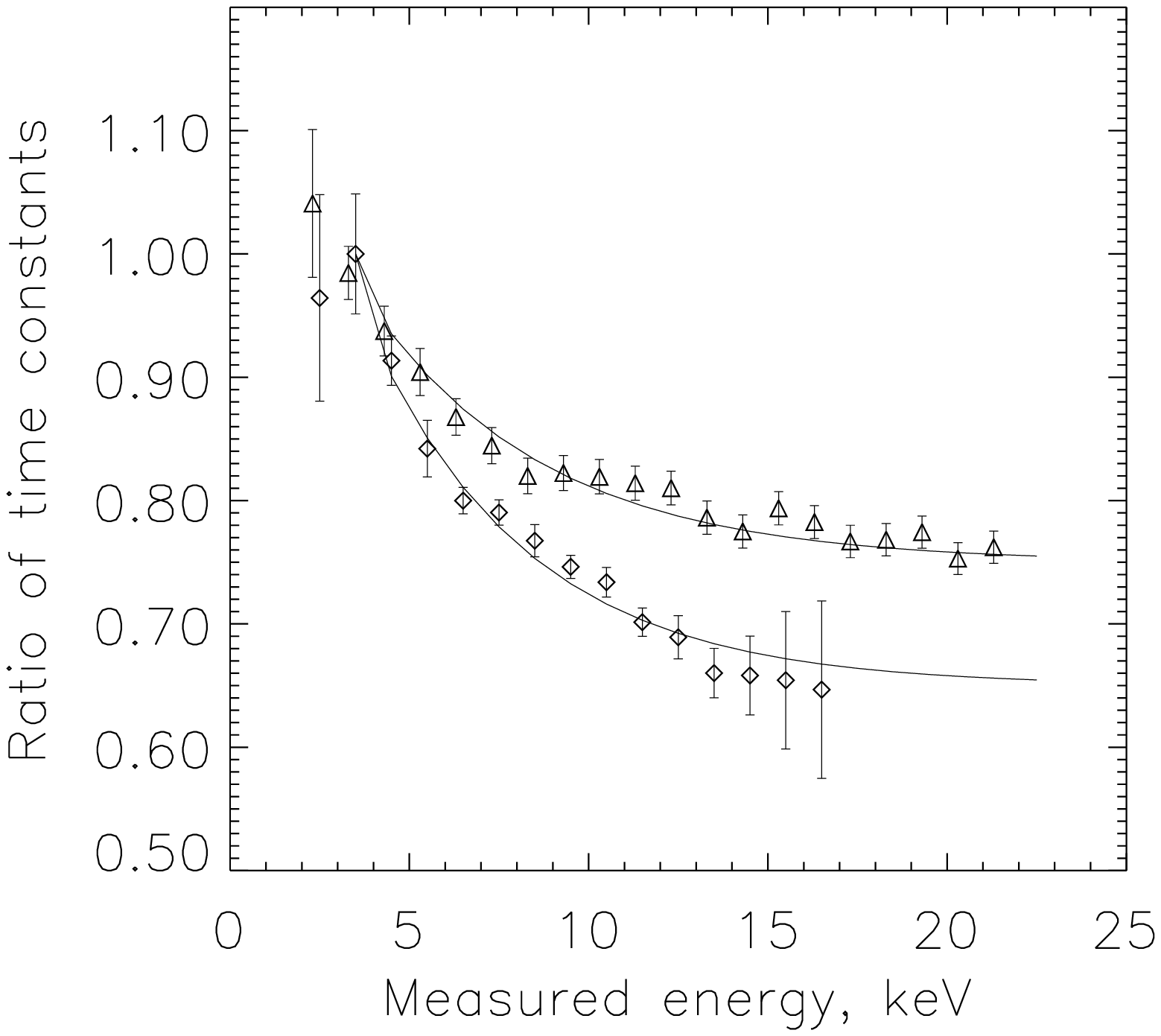,height=14cm}
\caption { }
\end{center}
\end{figure}

\pagebreak

\begin{figure}[htb]
\begin{center}
\epsfig{figure=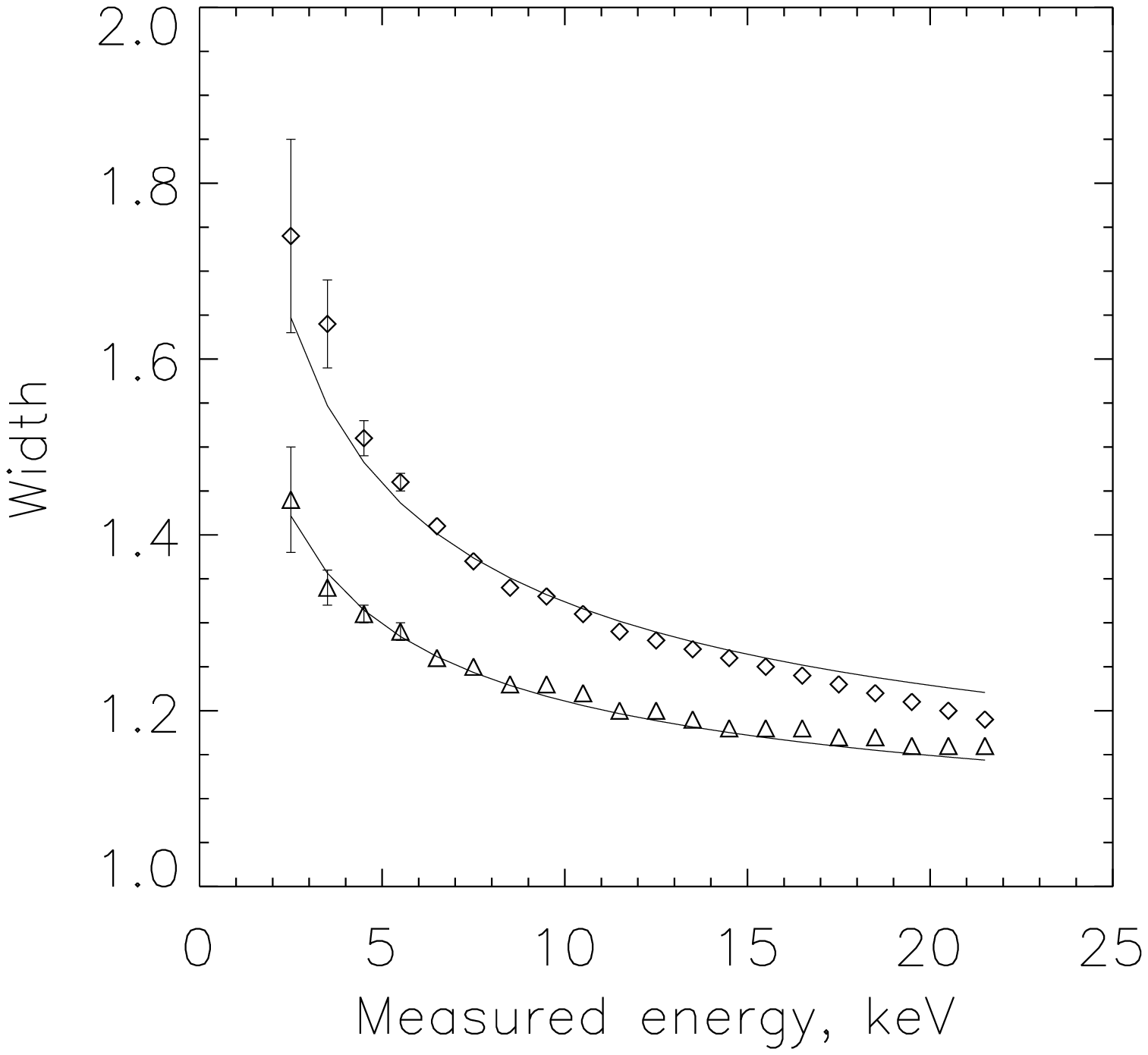,height=14cm}
\caption { }
\end{center}
\end{figure}

\pagebreak

\begin{figure}[htb]
\begin{center}
\epsfig{figure=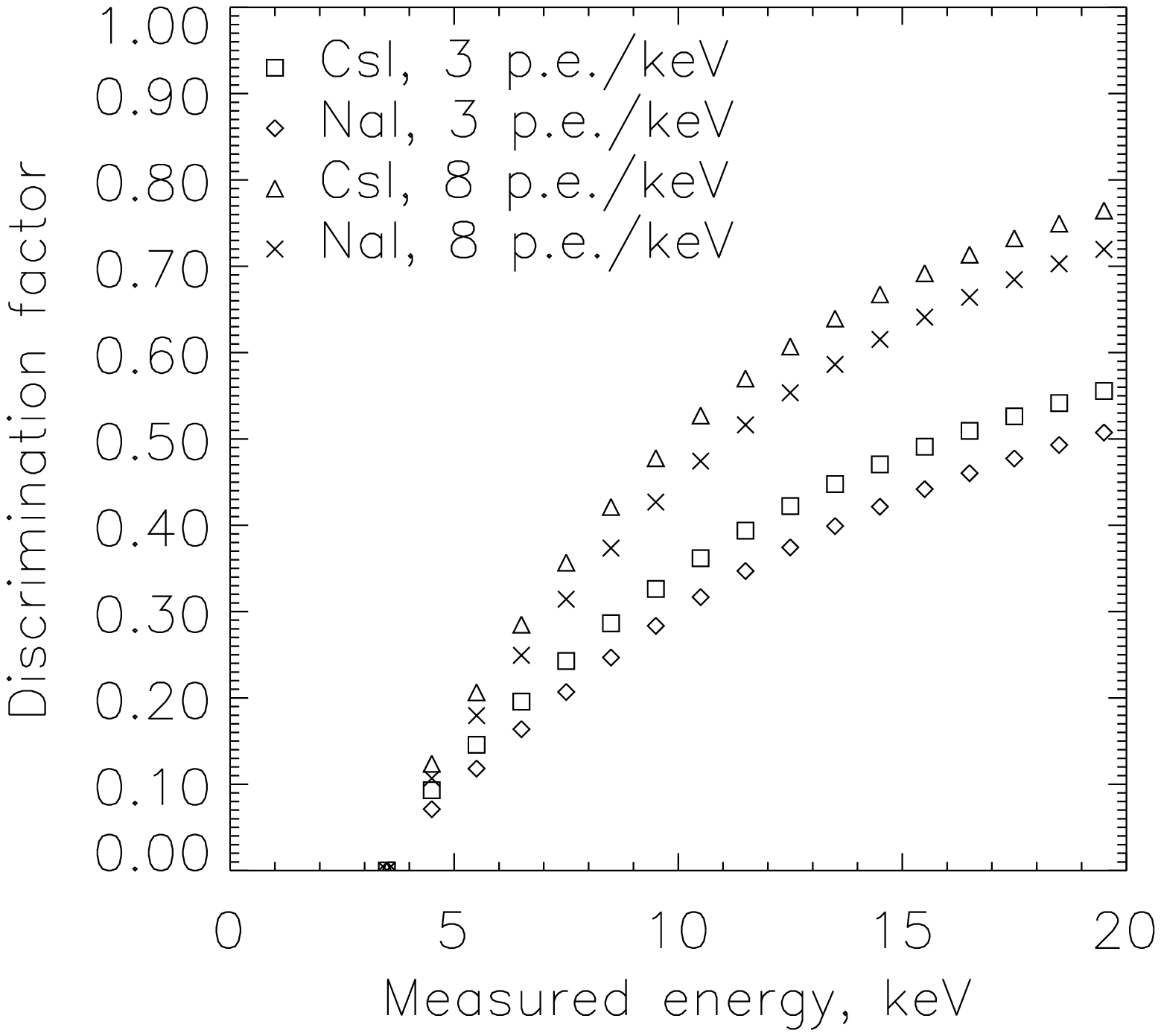,height=14cm}
\caption { }
\end{center}
\end{figure}

\pagebreak

\begin{figure}[htb]
\begin{center}
\epsfig{figure=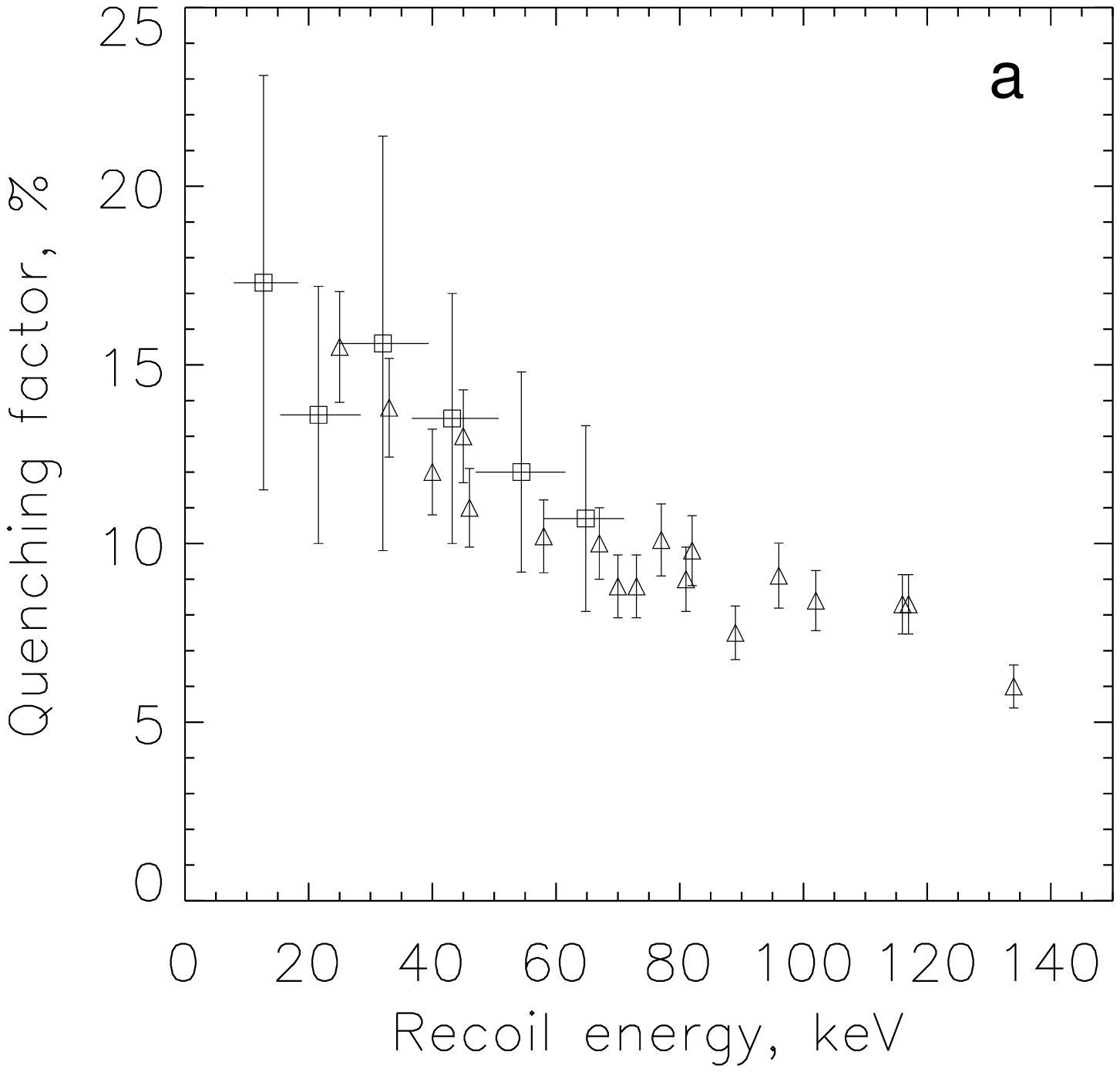,height=8cm}
\epsfig{figure=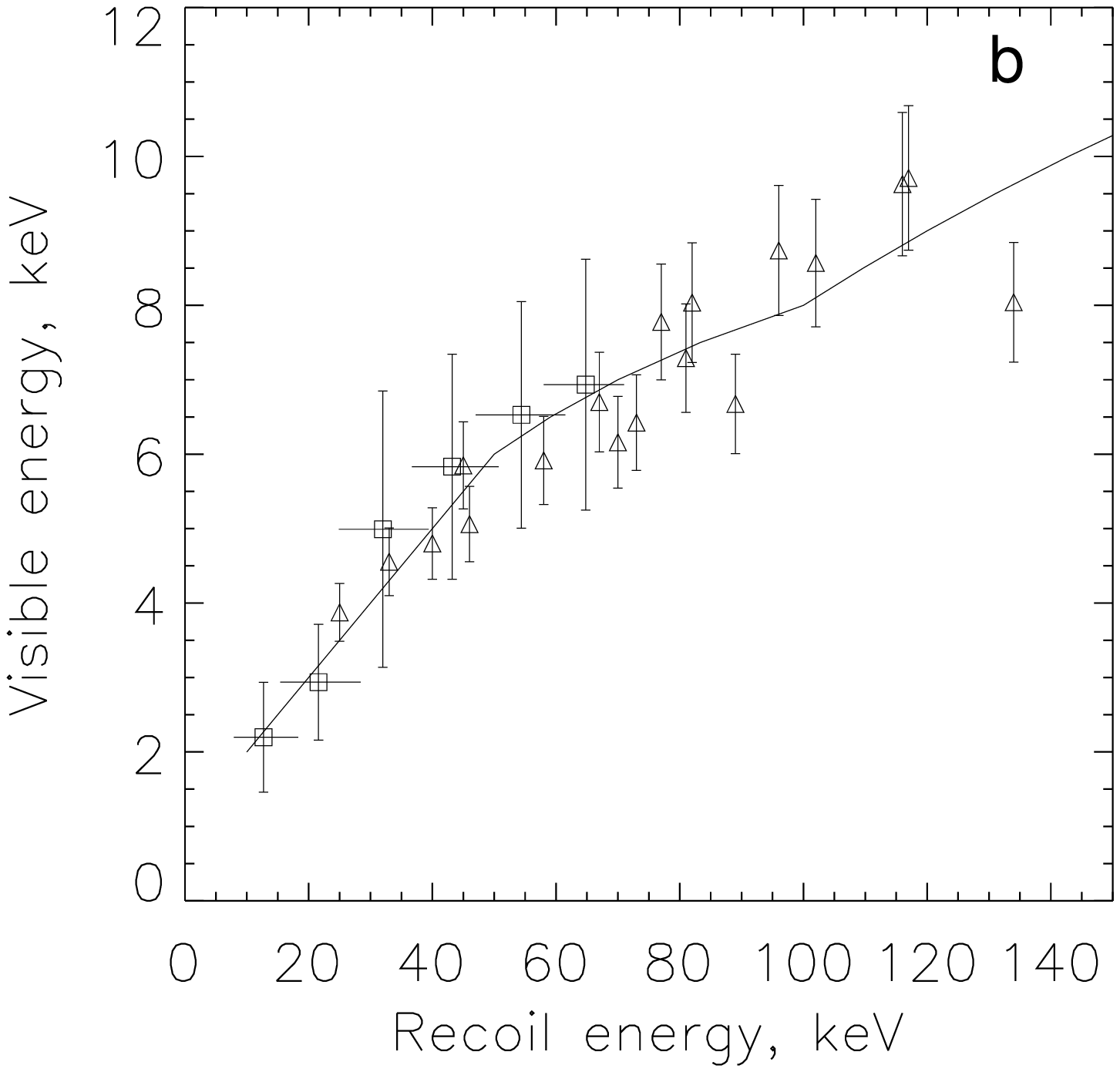,height=8cm}
\caption { }
\end{center}
\end{figure}
\pagebreak

\begin{figure}[htb]
\begin{center}
\epsfig{figure=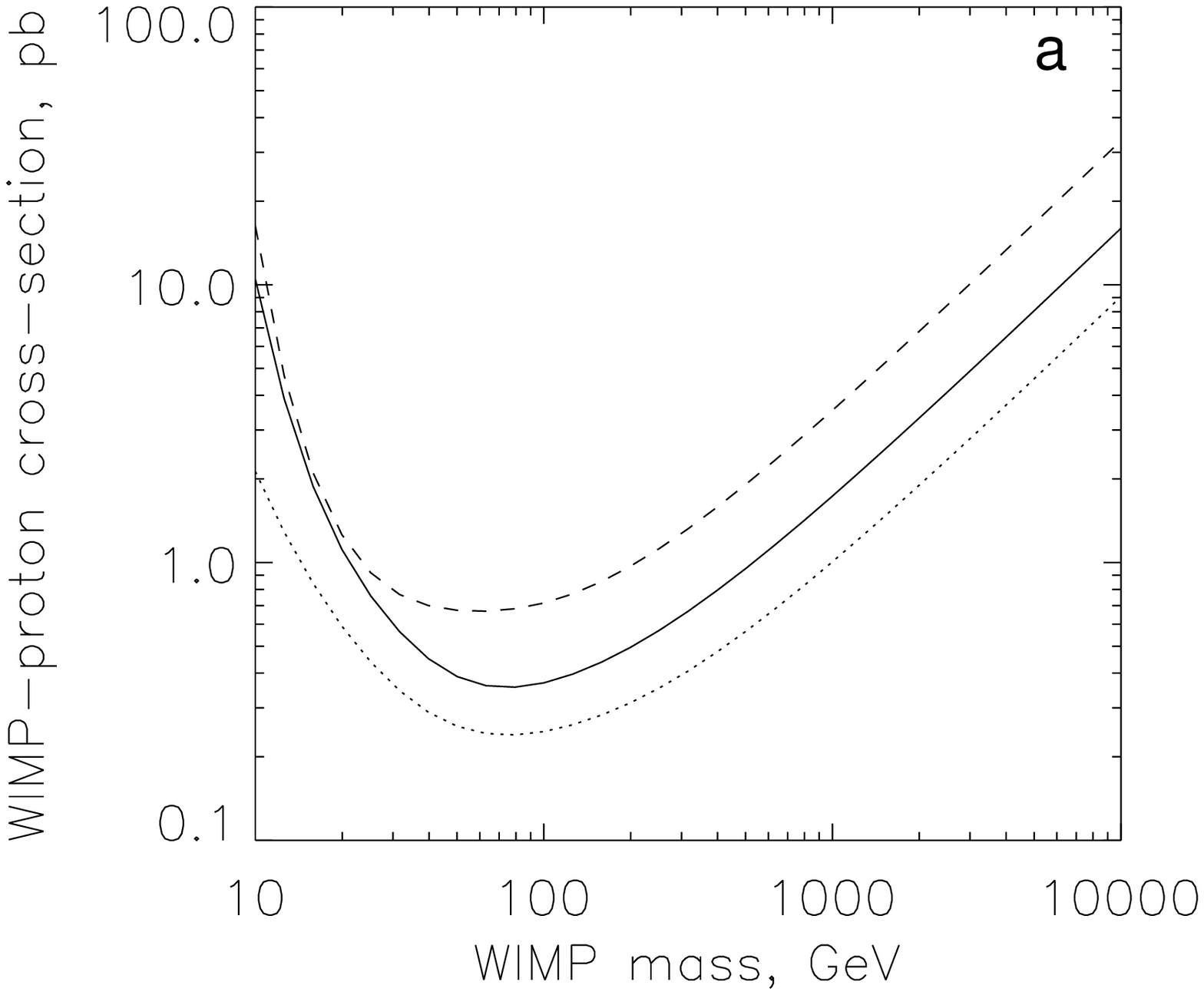,height=8cm}
\epsfig{figure=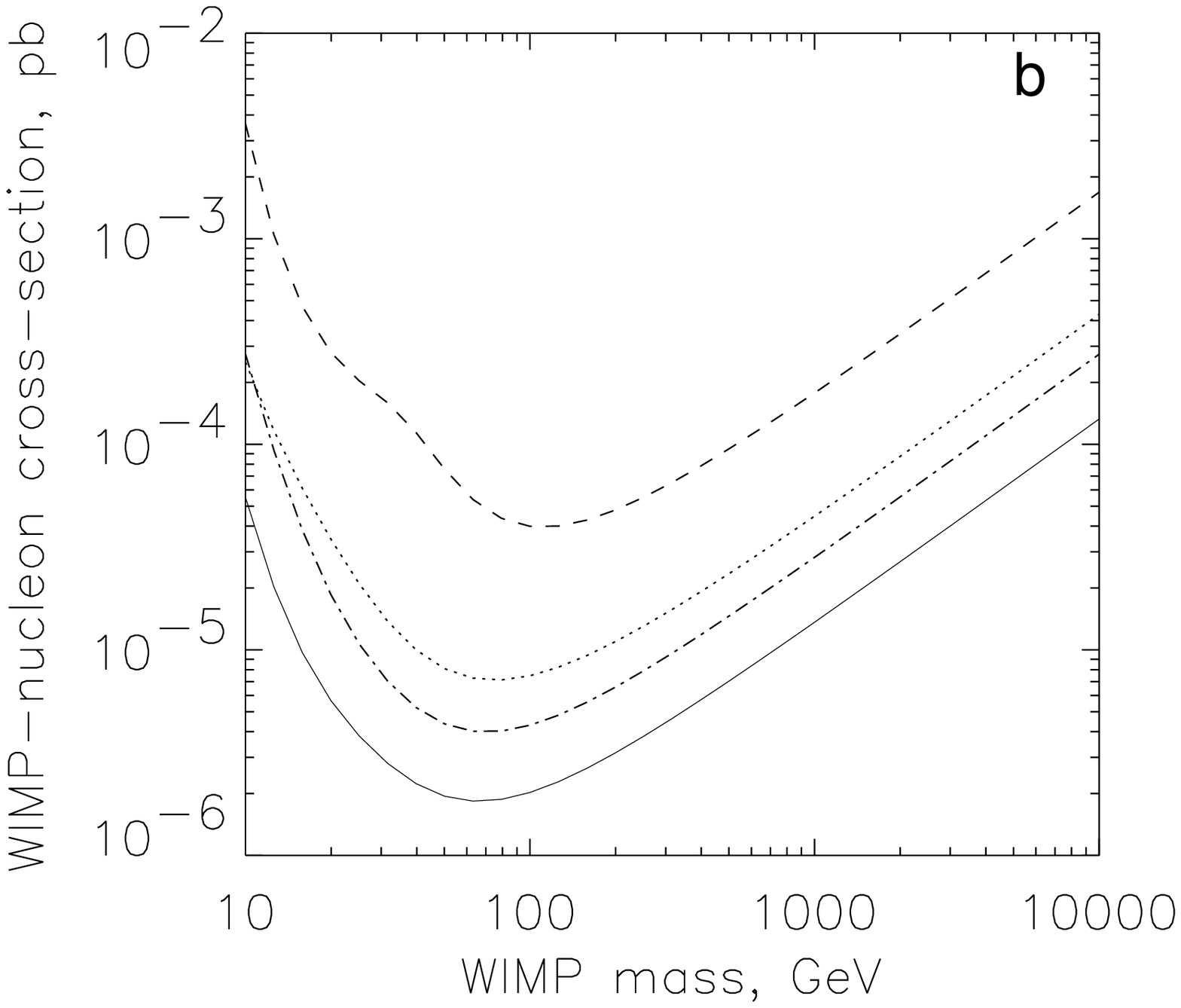,height=8cm}
\caption { }
\end{center}
\end{figure}

\end{document}